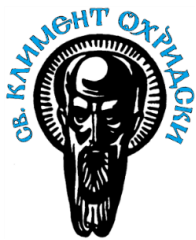

St. Kliment Ohridski University of Sofia

Department of Physical Chemistry, Faculty of Chemistry and Pharmacy

# Classical and Quantum Brownian Motion

Summary of the Dissertation

submitted for the Degree of Doctor of Science

by Prof. ROUMEN TSEKOV, M.Sc., Ph.D.

## CONTENTS

2021

**Introduction**

Methods of non-equilibrium statistical mechanics are the main tools for solving many theoretical problems of the contemporary physics and chemistry. Their aim is to describe time and spatial evolution of the macroscopic properties of matter on the base of the mechanical equations governing the underlying molecular dynamics. Historically this program has been realized in two ways: the kinetic theory of dilute gasses and the theory of Brownian motion. These two theories have played important role in the development of the statistical mechanics and are precursors of all modern statistical methods. The main achievement of the theory of dilute gasses is the famous Boltzmann equation, which is even nowadays the key for understanding of the behavior of many particles systems. Using it, all transport coefficients can be expressed by molecular characteristics. Unfortunately, finding the solution of the nonlinear integrodifferential Boltzmann equation is mathematically frustrated even for simple systems. Opposite to the kinetic theory of dilute gasses, the theory of Brownian motion operates with simpler mathematical apparatus and provides elegant solutions of many relaxation problems. It is the first attempt for stochastic modeling in science, which determines its importance not only for physics, chemistry and biology, but to mathematics as well. Many important methods for solving relaxation problems have originated from the Brownian motion, e.g. the theory of Markov random processes and the stochastic differential equations. Applications of the theory of Brownian motion are restricted, however, by two main limitations. First, it is a phenomenological theory, in the frames of which it is impossible to express the kinetic coefficients, governing the evolution of the non-equilibrium systems, via molecular parameters. Second, the theory of Brownian motion postulates the system tendency towards thermodynamic equilibrium with a known distribution, which a consistent non-equilibrium theory must derive by itself.

At present the Brownian motion is a synonym of thermal motion. The ways for its description are naturally divided in two groups: probabilistic and stochastic methods. The main feature of the probabilistic methods is replacement of the complex mechanical description of the interaction dynamics between the Brownian particle and the environment molecules via simple statistical hypothesizes about the random behavior of the Brownian particle. As a result, one yields linear differential equations for the evolution of various probability densities of the Brownian particle. Einstein is the founder of these methods for description of non-equilibrium processes in Nature. In his theory of Brownian motion, he introduced in 1905 a probabilistic diffusion equation for the first time in science and, in this way, initiated the successful story of the methods from this group. Nowadays the theory of Markov stochastic processes is the base of the probabilistic methods. Its main assumption is that the natural processes can be described well via a useful idealization without memory, called the Markov stochastic process. This type of stochastic processes is the second in the complexity hierarchy after those with independent events. The main feature of the Markov processes is that the entire information for the further evolution is given in the present state. Among the big diversity of Markov processes there is a class, known as diffusive Markov processes, whose trajectories are continuous, and for this reason they are widely used in the natural sciences. According to the theory of diffusive random processes, the probability density obeys the Fokker-Planck equation, which is the main tool for modeling of many kinetic phenomena in the modern science.

Attributing the Fokker-Planck equation to the Brownian motion yields many important results. For instance, the motion of a Brownian particle in the coordinate subspace is described by the Wiener process, while its behavior in the momentum subspace follows the Ornstein-Uhlenbeck process, providing the Maxwell distribution at equilibrium. The theory of diffusive processes describes also the Brownian motion in external potential fields via the Smoluchowski equation, which generates the Boltzmann distribution at equilibrium. Finally, the evolution of the probability density in the Brownian particle phase space obeys the Klein-Kramers equation, whose equilibrium solution is the Maxwell-Boltzmann distribution, as expected. There are exact methods for mapping the Klein-Kramers equation to another equation, governing the evolution of the probability density in the coordinate subspace only. The result is the telegraph equation, which possesses non-Markov character, but it reduces to the Smoluchowski equation in the case of adiabatic exclusion of the fast variables. As is seen, the probabilistic methods achieve many important results, which reflect the main features of the phenomenon. For this reason, every theory of the Brownian motion should reproduce them. The probabilistic methods suffer, however, some weaknesses as well. They are based on presumptions about the Brownian motion, which could be valid only in some peculiar cases. Various coefficients appear in the obtained kinetic equations but their physical meaning and estimation are outside the competence of the probabilistic methods. The latter are convenient for general description of the Brownian motion but cannot explain the underlining details and the elementary processes governing the phenomenon. These limitations require the usage of phenomenological constants and fluctuation-dissipation relations.

The main feature of the stochastic methods is introduction of a stochastic approximation of the Brownian particle dynamics. Thus, the derived stochastic differential equations can help in the calculation of the statistical properties of the Brownian motion and in the derivation of equations, analogical to the Fokker Planck equation, governing the evolution of the probability density. The founder of this kind of methods is Langevin, who has introduced in 1908 the first stochastic Newton equation for description of the Brownian dynamics. He has divided the interaction of the Brownian particle with the fluid molecules to two parts: a friction force, which in the frames of classical hydrodynamics is given by the Stokes formula, and a zero-centered random force, which accounts for the permanent kicks of the environment molecules on the Brownian particle. The solution of the Langevin equation is a particular stochastic realization of the Brownian particle trajectory. Using relatively simple assumptions about the statistical properties of the random force, one can obtain measurable statistical information for the Brownian motion. For instance, Langevin has proposed that the random force is not correlated to the position of the Brownian particle. Using this simple assumption and the equipartition theorem, relating the mean kinetic energy of the Brownian particle to the thermal one, Langevin has derived an expression for the dispersion of a free Brownian particle in the coordinate subspace, which is more general than the Einstein result. The latter follows asymptotically for times larger than the relaxation time, while in the opposite case the root-mean-square displacement equals to the thermal velocity multiplied by time, as expected. Moreover, the Langevin approach provides an expression for the friction coefficient even if macroscopic. In contrast to the probabilistic methods, it is possible now to derive all statistical characteristics of the Brownian motion, once the properties of the fluctuation force are known. The key model for the latter is the so-called white noise, which is a delta-correlated zero-centered stochastic process. Because it is equipartitioned sum of infinite number of Fourier components, the white noise is Gaussian due to the central limit theorem. The white noise is the first derivative of the Wiener process. By the science progress the Langevin ideas are further developed and enriched by concepts from the non-equilibrium statistical mechanics. Starting from general dynamic principles, Mori and Zwanzig have derived in 1965 a generalized Langevin equation, which basic innovation is the memory function, accounting for the past events on the present behavior of the Brownian particle. There are many models for the memory function, but it is rigorously shown that the Dirac delta-function memory describes a Brownian particle, which is much heavier than the environment molecules. In the case when the memory function differs from a delta-function, the generalized Langevin equation describes non-Markov random processes. A popular non-Markov model is the exponentially decaying memory, which is a particular example of a wide class of memory functions, generated by the Lee recurrence relation formalism.

Obviously, the stochastic methods describe better the Brownian motion, but their application requires special skills and erudition. As mentioned before, there are methods for derivation of the standard and generalized Fokker-Planck equations, starting from stochastic differential equations. The application of stochastic differential equations in physical chemistry is, however, accompanied by pitfalls and the most important difficulty is the correct distinction between the macroscopic value of a physical quantity and its fluctuations. Practically, many processes are described by the Langevin equations with a peculiarity in the integral form, due to the non-differentiable Wiener process. In contrast to the usual Riemann sums, which are independent from the selection of the middle point for integration, a stochastic integral sum depends essentially by our choice. In the modern science there are three standard alternatives for the middle point in the stochastic integrals: left end (Ito), center (Stratonovich) and right end (Hänggi-Klimontovich). From mathematical point of view all three forms are correct, but their applications in physical chemistry generate problems. The main difficulty is the correct physical splitting of the friction and diffusion in the Fokker-Planck equation, which are not independent, being related via the fluctuation-dissipation theorem. The advantage of the Ito choice is that the equilibrium value of the drift term is zero and, in this way, it is convenient for comparison with phenomenological equations. On the other hand, the Stratonovich choice respects the rules of traditional mathematics. The choice of Hänggi-Klimontovich is appropriate for the non-equilibrium thermodynamics since the drift term is proportional to the gradient of the equilibrium probability density. All three models are exact and the use of one of them is a matter of convenience.

Quantum mechanics was born in 1927 and it changed dramatically the notion of motion. Because the Schrödinger equation is a parabolic partial differential equation, its similarity to diffusion puzzled scientists from the inception of quantum mechanics. Indeed in 1966, Nelson succeeded to derive the Schrödinger equation from classical diffusion, but his derivation suffers a major shortcoming: the drift term in his stochastic differential equation depends on the probability density, which indicates a mean field approach. Moreover, it is known after Langevin that the trajectory of a classical Brownian particle is not differentiable only because to neglection of the inertial effects. Thus, the Wiener process is a mathematical approximation of the physical Brownian motion, while the status of the Schrödinger equation is exact. The latter is fundamental for physics and chemistry. It is proposed to reflect the controversial de Broglie idea that quantum particles can behave as waves as well, resembling the Einstein photons. Unfortunately, the Broglie relation, attributing a wavelength to every particle, misled generations of scientists to think about the electron as an object, which could be either particle or wave. The de Broglie wave-particle duality is unable, however, to answer many relevant physical questions. For instance, if an electron is a wave distributed in space, its negative charge must be divided somehow into many small pieces but such particles with infinitesimal fractional charges are never observed. Moreover, if the electron localizes back as a point particle, the related work against the electrostatic repulsion between the fractional charges would tend to infinity. In the second part of the Dissertation, we tried to convince the reader that electrons are particles, and they should experience quantum Brownian motion as well. Keeping the Schrödinger equation as experimentally proven, we explored another interpretation of quantum mechanics, where the particles remain points at any time as in classical mechanics. It is demonstrated that quantum mechanics is merely due to virtual force carriers transmitting the fundamental interactions. The force carriers are waves/quasiparticles in the coordinate/momentum subspaces, respectively, and this is the reason for the wavy character of quantum mechanics, not the point particles themselves. This physical picture is consistent with the quantum field theory.

    Despite the enormous progress of the modern quantum statistical physics, there are still problems in application of the developed concepts to complex systems. As a rule, the quantum theory of relaxation is less elaborated than the equilibrium one, but this is not surprising, since the same situation holds in classical physics as well. As was already mentioned, the theory of Markov random processes is the most powerful tool for description of the relaxation in classical systems. There are attempts for development similar quantum idealizations, for instance in the Glauber-Sudarshan or Husimi representations, but they are less general and universal. The most common approach to quantum dissipation is the division of a closed system to a subsystem and its environment. The isolated quantum system is rigorously described by the Schrödinger equation, which can be mathematically transformed to the Liouville-von Neumann equation, providing alternative description in terms of the more general density operator formalism. Integrating the Liouville-von Neumann equation over the environmental variables yields the master equation for the open quantum subsystem. The formal Nakajima-Zwanzig equation is the most general master equation, which reduces further to the Born-Markov equation in the case of weak subsystem-environment interactions and negligible memory effects. If additionally, the complete

positivity of the density matrix is required, one arrives to the Lindblad equation. All these equations are fundamentally linear but thermodynamic arguments point out that the exact master equation must be nonlinear. Indeed, the Schrödinger equation is linear for the wave function, while the classical Markov diffusion is linear for the probability density, which is the wave function square. Due to mathematical complications, this approach is applied to simple systems such as coupled harmonic oscillators. An alternative way to slide over the many particles difficulties is the use of dissipative Schrödinger equations, e.g. the Schrödinger-Langevin equation proposed by Kostin. Other methods are based either on extension of the Schrödinger equation to a stochastic differential one in the State Diffusion Theory or the Langevin-like description of the quantum dynamics in the frames of the time-dependent Heisenberg operators. In the third part of the dissertation a nonlinear master equation is derived, reflecting properly the entropy of the open quantum system. In contrast to the linear alternatives, its equilibrium solution is exactly the canonical Gibbs density matrix. The corresponding nonlinear equation for the Wigner function accounts rigorously for the thermo-quantum entropy and reduces at large friction to the Smoluchowski-Bohm equation in the coordinate subspace, which reflects the stochastic density-functional Bohm-Langevin dynamics. The Maxwell-Heisenberg relation for the momentum dispersion of quantum Brownian particles is discovered, which leads to the quantum generalization of the classical Einstein law of Brownian motion.

**1. Brownian Motion of Classical Particles**

The complete theoretical knowledge about the thermal motion of a mechanical subsystem of N particles in an arbitrary environment requires description of the dynamics of all atoms of the united system, interacting each other continuously. Because the subsystem atoms are heavy enough, their quantum nature will be ignored in this part of the dissertation. In the frames of classical mechanics, the mechanical definition of the entire system is governed by the Hamilton function $H$, which depends on the momenta and coordinates of all atoms. In a series of papers, we have considered a solid body as the thermal bath, where the bath particles can only vibrate near their equilibrium positions. If these oscillations are small, the Hamilton function can be modeled well in the frames of the harmonic approximation. In this case the bath particles are simple harmonic oscillators and substituting their classical trajectories in the coupled dynamic equations for the subsystem particles yields the generalized Langevin equation [2, 5]

$$M \cdot \ddot{X} + \int_0^t G(t,s) \cdot \dot{X}(s) ds = -\partial_X (U + \Phi) + F(X,t) \qquad (1.1)$$

Here $M$ is the diagonal 3Nx3N mass matrix, $X(t)$ is the 3N-dimensional vector of the trajectories of all subsystem particles, $U(X)$ is the interaction potential in the closed subsystem, and $\Phi(X)$ is the static interaction potential of the subsystem atoms with the bath particles, fixed at their equilibrium positions. As is seen, the stochastic Langevin force $F(X,t)$ is multiplicative and its dependence on $X$ goes through the forces, acting from the subsystem particles on the environment. The random behavior of the Langevin force originates solely from the unknown initial state of the bath particles. Employing the rigorous Gibbs distribution for the initial mechanical variables of the bath determines the statistical properties of the Langevin force. It is zero-centered and its autocorrelation function $C_{FF} = k_B T G(t,s)$ agrees with the Kubo second fluctuation-dissipation theorem for classical systems. Thus, the knowledge for the memory function $G$ is solely required in addition for the complete stochastic description by Eq. (1.1) of the mechanical subsystem coupled to the harmonic bath of the solid environment. The most general theory of the memory function is given in terms of the recurrence relation formalism developed by Lee. In the frames of the latter, we proposed a model, resembling the golden ratio, which was previously observed in the Brownian motion in liquids and fluctuating hydrodynamics [24].

The geometric Brownian motion is a particular example for a multiplicative Langevin force and it is applied for description of the financial markets since the beginning of the previous century. In fact, the Bachelier thesis "The Theory of Speculation", defended in 1900 under the supervision of Poincare, preceded the Einstein theory from 1905. In 1973, Black and Scholes have derived a famous formula for prizing of stock options, starting from the Fokker-Planck equation of the geometric Brownian motion. In general, the financial markets should be rigorously described via the exact generalized Langevin equation and we examined the golden ratio model, together with others, on the financial market dynamics [23]. Assuming Brownian self-similarity, the memory and autocorrelation functions of the market return rate are derived, which exhibit oscillatory-decaying behavior with a long-time tail, similar to many empirical observations. It is responsible in particular for the well-known Elliott waves of the market prizes. Individual stocks are also described via the generalized Langevin equation. They are classified by their relation to the market memory as heavy, neutral, and light stocks, possessing different kinds of autocorrelation functions.

Returning to physical chemistry, the memory function is determined by all vibrations in the solid thermal bath. Usually, the acoustic phonons in solids are modeled well by the Debye spectrum. Since the Debye frequency, being the upper cutoff, is huge, the corresponding memory function approximates well by to the Dirac delta-function, which is typical for the Markov processes. Substituting the corresponding $G = 2B(X)\delta(t-s)$ in Eq. (1.1), the generalized Langevin equation reduces to an ordinary Langevin equation for the subsystem particles [2, 5]

$$M \cdot \ddot{X} + B(X) \cdot \dot{X} = -\partial_X (U + \Phi) + [2k_B T B(X)]^{1/2} \cdot F(t) \tag{1.2}$$

where $F(t)$ is a white noise. The last expression for $F(X,t)$ must be understood as the Hänggi-Klimontovich product [8]. An important feature of Eq. (1.2) is the position-dependent friction tensor $B$. Presuming short-range interactions between the subsystem and environment atoms, we derived a very important relation $(4\pi\rho_m c^3 B)^{1/2} = \partial_x \partial_x \Phi$, linking the dissipative friction tensor to the conservative subsystem-bath interaction potential $\Phi$, where $\rho_m$ and $c$ are the mass density and sound velocity of the solid, respectively, originating from the Debye frequency. To examine this relation, let us consider some simple examples. The small vibrations of an atom, adsorbed on the solid surface, corresponds to the harmonic interaction potential $\Phi = m\omega_0^2 x^2 / 2$ and the relevant friction constant $b = m^2 \omega_0^4 / 4\pi\rho_m c^3$ depends strongly on the oscillator own frequency $\omega_0$. If the barrier potential $\Phi = -m\omega_1^2 x^2 / 2$ is considered at desorption, the friction constant is similar but depends on the barrier frequency $\omega_1$ now. In the case of surface diffusion, the periodic Frenkel-Kantorova model $\Phi = A\cos(2\pi x / a)$ yields a friction coefficient $b = 4\pi^3 \Phi^2 / \rho_m c^3 a^4$ which is also periodic. The latter is not constant, which is important, because the scientific literature is crowded by publications, describing Brownian dynamics in structured media with constant friction coefficients, which is a very rough approximation. We extended Eq. (1.2) to amorphous solids, which are frozen liquids [5]. Exploring the idea for the so-called instantaneous normal modes, the model is extended to usual liquids as well via 'melting' the amorphous solid over the Gibbs quasi-equilibrium configurations. Thus, the static potential $\Phi$ is replaced in Eq. (1.2)

via the bath conditional free energy, which particularly for N=1 is uniform, and the friction coefficient does not depend anymore on the position of the lonely Brownian particle, as expected.

In classical physics, the probability density of diffusive Markov processes obeys the linear Fokker-Planck equation. A particular example is the Klein-Kramers equation

$$\partial_t f + \partial_p H \cdot \partial_x f - \partial_x H \cdot \partial_p f = \partial_p \cdot B \cdot (f \partial_p H + k_B T \partial_p f) \quad (1.3)$$

which describes the evolution of the probability density $f(p,x,t)$ in the phase space of all momenta and coordinates of the subsystem particles. The Hamilton function, corresponding to the Langevin equation (1.2), reads $H = p \cdot M^{-1} \cdot p / 2 + U + \Phi$. If $B$ is annulled, Eq. (1.3) reduces to the Liouville equation, being equivalent to classical mechanics of the closed subsystem. Furthermore, the special relativity can be also treated by Eq. (1.3) via the relevant Einstein expression for the Hamilton function. The relaxation term on the right-hand side drives the irreversible evolution towards the thermodynamic equilibrium. The equilibrium solution of Eq. (1.3) is the canonical Gibbs distribution $f_{eq} = \exp(-\beta H)/Z$, where $\beta \equiv 1/k_B T$ is the reciprocal temperature. The partition function $Z$ determines the equilibrium free energy $F_{eq} \equiv -k_B T \ln Z = H + k_B T \ln f_{eq}$ which is the characteristic potential of the subsystem and contains the entire thermodynamic information. Thus, any problem in classical statistical mechanics and thermodynamics could be solved via Eq. (1.3), in principle, once the mechanical definition is specified by $H$.

For many applications it is sufficient to describe only the evolution of the probability density $\rho(x,t) \equiv \int f dp$ in the coordinate subspace. Integrating properly Eq. (1.3) yields the corresponding evolutionary Euler-like equations

$$\partial_t \rho + \partial_x \cdot (\rho V) = 0 \qquad M \cdot (\partial_t V + V \cdot \partial_x V) + B \cdot V = -\partial_x (U + \Phi + k_B T \ln \rho) \quad (1.4)$$

The first equation is the continuity equation, following from the compulsory conservation of the probability, and $V(x,t)$ is the hydrodynamic-like velocity in the probability coordinate subspace. The second equation is the macroscopic force balance resembling the Langevin equation (1.2). One can recognize the configurational entropy in the last logarithmic term, whose gradient is the macroscopic image of the stochastic Langevin force. If the friction is high, one can neglect the inertial term as compared to the friction force to get $V = -B^{-1} \cdot \partial_x (U + \Phi + k_B T \ln \rho)$. Substituting this expression for the hydrodynamic-like velocity into the continuity equation yields the self-consistent Smoluchowski equation

$$\partial_t \rho = \partial_x \cdot B^{-1} \cdot [\rho \partial_x (U + \Phi) + k_B T \partial_x \rho] \tag{1.5}$$

It is evident that the equilibrium solution of the Smoluchowski equation is the Boltzmann distribution, which is not disturbed by the non-uniformity of the friction tensor $B$. The latter affects solely the evolution towards equilibrium but not the final equilibrium state.

The Brownian motion through modulated structures is important process either for the academic science or for technology. As a rule, the diffusion coefficient of molecules exhibits non-monotone dependence of their size, a phenomenon called resonant diffusion. For example, Gorring has experimentally detected in 1973 that the diffusion coefficient of normal alkanes in zeolites exhibits a minimum at $C_8$ and maxima at $C_4$ and $C_{12}$. The resonant Brownian motion of small clusters on solid surfaces is important for catalysis, heterogeneous nucleation, surface coating, etc. In biology the transport of molecules through the highly structured biological membranes is another example, where the resonant diffusion takes place. Our aim now is to derive qualitative description of the resonant diffusion, which is very general and could be applied to diffusion in any modulated structure. The basic assumptions are that the structural vibrations are harmonic, and the diffusion mechanics is interstitial, while the effect of vacancies is of second order of importance. Because of the large difference between the characteristic times of diffusion and molecular vibrations, the adiabatic separation of slow and fast variables is possible, and the latter can be considered as degrees of freedom at equilibrium. In 1978, Festa and d'Agliano solved Eq. (1.3) for a single Brownian particle. Their result $D = k_B T / [\overline{b \exp(\beta \Phi)} \, \overline{\exp(-\beta \Phi)}]$ for the diffusion constant accounts for the position dependence of the friction coefficient $b$ and the potential $\Phi$ by means of the spatial average, indicated via bars. Considering the potential as a strictly periodic with a spatial period $a$, the Festa-d'Agliano formula reduces to the more specific Lifson-Jackson form, written here by employing our expression for the friction coefficient [3, 4, 9],

$$D = 4\pi \rho_m c^3 k_B T a^2 / [\int_0^a (\partial_x^2 \Phi)^2 \exp(\beta \Phi) dx \int_0^a \exp(-\beta \Phi) dx] \tag{1.6}$$

which requires solely a model for the static subsystem-bath interaction potential. Since at low temperature the two integrals are essential near the maxima and minima of the potential $\Phi$, respectively, the diffusion coefficient (1.6) decreases exponentially with the height of the energy barriers, which premises the Arrhenius law. Note that $b$ takes place in the first integral only, i.e. the position dependent friction is more important at the barrier hills than at the potential wells. For a diffusing atom in the Frenkel-Kantorova model Eq. (1.6) can be explicitly integrated and the corresponding diffusion coefficient $D = \rho_m c^3 a^4 / 4\pi^3 A[\beta A I_0(\beta A) - I_1(\beta A)] I_0(\beta A)$ is expressed via the modified Bessel functions of the first kind [4]. Since the latter tend to $\exp(\beta A)/(2\pi\beta A)^{1/2}$ at large argument, the diffusion coefficient $D = (\rho_m c^3 a^4 / 2\pi^2 A)\exp(-2\beta A)$ recovers the Arrhenius law with a pre-exponential factor, decreasing by the activation energy increase. In the opposite case at high temperature, the diffusion coefficient reduces to the Einstein formula $D = k_B T / \bar{b}$ with the spatially averaged friction constant $\bar{b} = 2\pi^3 A^2 / \rho_m c^3 a^4$. Note that even if the hopping over the energetic barriers is easy, the diffusion coefficient decreases further by the increase of the interaction strength $A$ between the Brownian particle and the environment, due to the increase of the friction.

When the Brownian particle is a dimer, moving in one direction along a surface, the Frenkel-Kantorova periodic surface potential $\Phi = 2A\cos(\pi l \cos\phi / a)\cos(2\pi x / a)$ depends also on the dimer length $l$ and the angle $\phi$, enclosed between the direction of motion and the dimer. It premises resonant diffusion since the interaction strength becomes even zero at some geometric conditions. Employing this potential, we have described several particular examples of rigid [4], rotating [6] and vibrating [12] dimers on solid surfaces and our results correlate well to existing experimental observations. In all cases, the dimer internal degrees of freedom ($A, l, \phi$) are averaged properly via the adiabatic quasi-equilibrium conditional Gibbs probability density. The model is further developed to describe linear chains of N interacting Brownian particles to explain the Gorring observation [3, 9]. The diffusion of particles through zeolites is a very important mass-transfer process, which is related to the sorption kinetics and catalytic activity. The regular pore structure of zeolites provides desired selectivity by the access of the diffusing molecules to the catalytic centers. The 3D Brownian motion of chains in the zeolite structure is very complex. To reduce the system to the Frenkel-Kantorova model, we selected data only from linear zeolites, such as ZSM-12 and LTL, where the Brownian motion is restricted in a single channel in one direction. As was mentioned, an interesting aspect of the diffusion in zeolites is the resonant effect.

Because of the shape of the channels, the resultant interaction potential is not monotone function from the geometric parameters of the Brownian particle and the zeolite. In this way, it could happen that longer alkane chains possess lower energy barriers than shorter chains, which will result in faster diffusion as observed experimentally by Gorring.

The specific interaction parameters per a methylene group could be estimated from the data for diffusion of methane in the considered zeolite. In general, the structural periodicity of the zeolite crystals can be approximated by two Fourier components. The short length periodicity reflects the atomic structure of the zeolite. Since the lengths of the Si-O and C-C chemical bonds are commensurable, one can accept $l$ as the short period of the zeolite crystal as well. The long-range periodicity is due to the zeolite pore structure. Because the friction constant $b$ is proportional to the square of the second derivative of the interaction potential, it follows that the friction coefficient is determined mostly by the atomic structure of the zeolite and increases linearly with the increase of the number of C-atoms in the alkane. On the contrary, the effective interaction potential of the diffusing molecule depends essentially on the pore structure [9]. The calculated diffusion constant in LTL, for instance, exhibits periodic resonant character with maxima for $C_6$, $C_{12}$, $C_{18}$ and minima for $C_3$, $C_9$, $C_{15}$, which is also confirmed by our experimental observations [11]. The detail description of the stochastic dynamics allows to determine the effect of some hidden degrees of freedom. It is confirmed, for instance, that the diffusion coefficient increases with the stiffness of the chain, which is opposite of the normally expected easier movement of flexible molecules. This phenomenon is explained by disturbance of the zeolite phonon density by the vibrating molecular bonds, reflecting an increase of the activation energy.

The Langevin equation (1.2) can be extended to describe quantum effects in the environment as well. Considering the bath oscillators as quantum, the Langevin force $F(t)$ becomes the quantum noise. Its spectral density $S_{FF} = (\beta\hbar\omega/2)\coth(\beta\hbar\omega/2)I$ follows from the quantum fluctuation-dissipation theorem where $I$ is the 3Nx3N dimensional unit tensor. It resembles the well-known expression for the thermal phonon energy and recovers logically the spectral density of the white noise $S_{FF} = I$ in the classical limit $\hbar \to 0$. Inverting the Fourier image $S_{FF}$ yields the corresponding Langevin force autocorrelation function, which can be written conveniently in the formal operator form $C_{FF} = (\beta i\hbar\partial_t/2)\coth(\beta i\hbar\partial_t/2)\delta(t-s)I$. For compactness, we will employ further the corresponding temperature operator [30]

$$k_B \hat{T} \equiv \coth(\beta \hat{E}/2)\hat{E}/2 = \cot(\beta \hbar \partial_t / 2)\hbar \partial_t / 2 \qquad (1.7)$$

which is defined via the quantum operator of energy $\hat{E} \equiv i\hbar \partial_t$. The latter emphasizes the origin of the quantum effects because the quantum bath supplies simply energy to the classical subsystem in a more complex way. The temperature operator $\hat{T}$ reduces naturally to temperature $T$ in the classical limit, while expanding it in series $k_B \hat{T} = k_B T - \beta \hbar^2 \partial_t^2 / 12$ yields the leading semi-classical correction. Since the quantum Langevin force is Gaussian, the relevant quantum Klein-Kramers and Smoluchowski equations follow directly by application of the Furutsu-Novikov-Donsker theorem [30]. Therefore, Eq. (1.3) and Eq. (1.5) retain their structures by replacing solely the temperature $T$ via the temperature operator $\hat{T}$. These equations describe non-Markov processes, because they involve time-derivatives higher than the first one. The equilibrium solutions of the quantum Klein-Kramers and Smoluchowski equations do not depend on time and, hence, they are the classical Gibbs and Boltzmann distributions, respectively. This is expected, since the equilibrium follows rigorously at infinite time, which corresponds to zero frequency, where the quantum Langevin force behaves classically. Therefore, the quantum bath modulates exclusively the evolution of the classical subsystem to its classical equilibrium state.

The stochastic description by Eq. (1.2) is frustrated by several nonlinearities. Considering a free Brownian particle ($U \equiv 0$), one can simplify the Langevin equation further by employing of the spatially averaged friction tensor $\bar{B} = \bar{b}I$, which is diagonal. Obviously, $\bar{b}$ is independent of the particle position and $\bar{\Phi} = 0$ is constant. These approximations significantly reduce the mathematical complexity and make possible to use the standard Fourier analysis. Thus, from the equilibrium spectral density of the Brownian particle velocity $S_{\dot{X}\dot{X}} = \bar{b}\hbar\omega \coth(\beta\hbar\omega/2)/(m^2\omega^2 + \bar{b}^2)$ one can derive at zero frequency the Einstein diffusion coefficient $D = S_{\dot{X}\dot{X}}(\omega = 0)/2 = k_B T / \bar{b}$, which is also classical. The integral of the spectral density $S_{\dot{X}\dot{X}}$ on $\omega$ should provide the equilibrium thermal dispersion $k_B T / m$ and this determines the maximal frequency of the spectrum, $\Omega \approx (2\pi \bar{b} k_B T / m\hbar)^{1/2}$ [30]. As is seen, it diverges in the classical limit, because all Fourier modes are present in the classical bath with a white noise. Interpreting $\Omega$ as the maximal collision frequency suggests that the smallest mean free path of the Brownian particle is restricted by the quantum environment. It scales with $(\hbar/\bar{b})^{1/2}$, which appears often in the theory of Brownian motion in quantum environments. Expressing now the friction coefficient $\bar{b} = \hbar/\lambda^2$ via the mean free path $\lambda$ yields an example of quantum friction in gases [21].

Finally, we are going to conclude the first part of the Dissertation by describing a stochastic phenomenon from the living world, where the migration of living cells usually obeys the laws of Brownian motion as well. While the latter is due to thermal fluctuations in the environment, the locomotion of cells is generally associated with their vitality and active swimming. We studied theoretically the driving force of cell migration and proposed a model for the Brownian motion of cells [22]. Accordingly, another effective temperature appears as the main parameter, which we called the cell temperament $\theta$. The latter is a biophysical parameter describing the motivity of living biological entities in analogy with the physical parameter of temperature, which dictates the movement of lifeless physical objects. In this respect, an interesting question arises here: what is the osmotic pressure of fishes in an aquarium? If the cell is death the temperament $\theta$ coincides with the thermal energy $k_B T$. The cell migration is studied via the generalized Langevin equation (1.1). We explored the possibility to describe the cell locomemory via the Brownian self-similarity [24]. In the frames of the Markov approximation, the corresponding cell Klein-Kramers and Smoluchowski equations are derived, where temperature is replaced by the temperament $\theta$. Thus, the equilibrium Maxwell-Boltzmann distribution also describes the probability density for the velocity and position of living cells. A heuristic expression for the diffusion coefficient of cells on structured surfaces is proposed by exploring the analogy to the Festa-d'Agliano formula (1.6).

## 2. Brownian Motion and Quantum Mechanics

A century after the inception of quantum mechanics, there are still attempts to derive it from the Brownian motion. The reasoning is due to the a priori probabilistic character of quantum mechanics and the parabolic form of the Schrödinger equation, resembling diffusion with an imaginary diffusion coefficient [28]. Since the latter describes a mechanical subsystem in vacuum, there are not bath atoms and the generalized Langevin equation (1.1) reduces at $\Phi \equiv 0$ to the Newton equation $M \cdot \ddot{X} = -\partial_X U$ of the closed subsystem. However, the lack of atoms does not mean that vacuum is empty. The quantum field theory states that there are virtual photons in vacuum, carrying the zero-point energy fluctuations. Considering a single electron, there is another way for energy dissipation via emission of electromagnetic radiation. The ability of quantum photons to travel freely in vacuum makes the latter a very dissipative environment. In a first approximation, the radiation friction can be described via the Abraham-Lorentz force, which must be compensated by the quantum Langevin force, originating from the zero-point energy fluctuations [25]. Thus, one arrives to $m\ddot{X} - e^2 \dddot{X}/6\pi\varepsilon_0 c^3 = -\partial_X U + F$ and this generalized Langevin equation describes a non-Markov process, because the friction force is linear on the electron jerk. The spectral density the fluctuation force $S_{FF} = (e^2\omega^2/6\pi\varepsilon_0 c^3)\hbar\omega$ follows from the quantum fluctuation-dissipation theorem at zero temperature. As is seen, the quantum Langevin force is not a white noise, and it is relativistic as well. It vanishes at infinite speed of light $c$, which replaces the speed of sound in the friction coefficient. Because the generalized Langevin equation above becomes linear at $U \equiv 0$, let us consider a free electron. Applying the standard Fourier analysis to this generalized Langevin equation yields the equilibrium spectral density of the electron velocity $S_{\dot{X}\dot{X}} = \hbar\omega\tau_0/m(\omega^2\tau_0^2 + 1)$, where the characteristic time $\tau_0 \equiv e^2/6\pi\varepsilon_0 mc^3$ of photon emission by the moving electron is extremely small, being of the order of a yocto-second. Because of $S_{\dot{X}\dot{X}}(\omega=0) = 0$, the classical diffusion coefficient is zero as well, which is related to the zero temperature in vacuum. The integral of $S_{\dot{X}\dot{X}}$ on $\omega$ diverges in general, which points out the existence of a maximal frequency as the Debye cutoff, which is the Zitterbewegung frequency $\Omega = 2mc^2/\hbar$ now. Since the product $\Omega\tau_0 = e^2/3\pi\varepsilon_0\hbar c$ is of the order of the small fine structure constant, the velocity spectral density simplifies further to $S_{\dot{X}\dot{X}} = \hbar\omega\tau_0/m$, being directly proportional to the zero-point energy $\hbar\omega/2$. Integrating this expression yields the velocity dispersion of the electron in vacuum at $T = 0$: $\sigma_{\dot{X}}^2 = e^2 c/3\pi^2\varepsilon_0\hbar$ is universal, relativistic and quantum. Hence, the speed of the trembling motion of a free electron in vacuum is an order slower than the speed of light $c$ and an order faster than the orbital electron in the hydrogen atom. The relativistic quantum model described above is the core of the stochastic electrodynamics, which is considering it as the background of quantum mechanics. However, instead of the Schrödinger equation the Brownian emitter model leads to the Klein-Kramers and Smoluchowski equations, generalized by the semi-relativistic friction coefficient operator $\hat{b} = -e^2\partial_t^2/6\pi\varepsilon_0 c^3$ [25, 30]. Because the latter is acting in time as the temperature operator $\hat{T}$, they do not affect the equilibrium solutions. Furthermore, the Schrödinger equation is valid for non-charged particles as well, which cannot emit electromagnetic radiation. The Brownian model considered here points out, however, the fast-stochastic irregular relativistic trembling motion as the possible origin of quantum mechanics.

Because quantum mechanics is a probabilistic theory, it is more natural to reformulate it in terms of the hydrodynamic-like equations for the probability density, analogical to Eq. (1.4). In 1927, Madelung has already converted the Schrödinger equation for a single particle to the hydrodynamic Euler equations. Employing the Madelung ansatz $\psi = \rho^{1/2} \exp(iS/\hbar)$ for the wave function, which reflects the Born rule as well, the Schrödinger equation transforms to the following hydrodynamic-like equations [1, 19, 29]

$$\partial_t \rho + \partial_x \cdot (\rho V) = 0 \qquad M \cdot (\partial_t V + V \cdot \partial_x V) = -\partial_x (U + Q) \qquad (2.1)$$

where the hydrodynamic-like velocity $V \equiv \hbar M^{-1} \cdot \partial_x S$ is related to the wave function phase $S$. As is seen, the Bohm quantum potential $Q \equiv -\hbar^2 \rho^{-1/2} \partial_x \cdot M^{-1} \cdot \partial_x \rho^{1/2}/2$ accounts for all quantum effects alone. Comparing Eq. (2.1) and Eq. (1.4) reveals some peculiarities of the quantum vacuum as environment. As expected, there is no interaction potential $\Phi$ and the lack of friction force hints against the radiation mechanism. On the other hand, the thermal free energy is replaced by the quantum potential in Eq. (2.1). Obviously, $Q$ possesses an emergent origin [26], since the quantum potential is a functional of the probability density $\rho$, in contrast to the usual potentials. Since the average value of the quantum potential is proportional to the Fisher information, $Q$ is sometimes referred as the information potential as well. On the other hand, the classical entropy is related to the Shannon information. Thermodynamically, at zero temperature the quantum potential is a kind of subsystem enthalpy in vacuum, being the characteristic potential at constant entropy. Thus, its gradient $\partial_x Q = \partial_x \cdot \Pi / \rho$ determines the Madelung pressure tensor $\Pi = -\hbar^2 \rho M^{-1} \cdot \partial_x \partial_x \ln \rho / 4$, which is the quantum analog of the thermal osmotic pressure tensor $k_B T \rho I$ from Eq. (1.4).

Looking for the underlining stochastic dynamics behind Eq. (2.1), it is natural to introduce stochastic Euler equations in the probability space, driven by the zero-point vacuum fluctuations [29, 32]. The fast and irregular relativistic trembling of the quantum particles can cause turbulence in the hydrodynamic-like stochastic equations. Thus, the turbulent character of the Madelung pressure tensor is elucidated by splitting $\Pi = pI + \rho M^{1/2} \cdot V' M^{1/2} \cdot V'$ to a scalar pressure and the Reynolds stress tensor. The quantum pressure $p = -\hbar^2 \partial_x \cdot M^{-1} \cdot \partial_x \rho / 4$ is given by the kinetic energy operator acting on the probability density. Using it, one can calculate the local-mean quantum force $-\partial_x p / \rho$ in Eq. (2.1), being the macroscopic image of the quantum Langevin force [27]. The Reynolds stress tensor is dyadic and $V' \equiv -\hbar M^{-1} \cdot \partial_x \ln \rho / 2$ represents the local-mean turbulent velocity. Remarkably, the flow $\rho V' = -\hbar M^{-1} \cdot \partial_x \rho / 2$ obeys the Fick law with the Nelson quantum diffusion tensor $\hbar M^{-1} / 2$, which confirms that the quantum turbulence possesses a Brownian character. Indeed, using the quantum momentum operator $\hat{p} \equiv -i\hbar \partial_x$ and the Madelung ansatz for the wave function one can write $\hat{p}\psi = M \cdot (V + iV')\psi$. Because $V$ is the local-mean laminar velocity of the quantum particles probability density flow, it appears that the kinetic energy from the Schrödinger equation is a sum of the energy contributions of the laminar and turbulent motions. This is again an illustration for separation to slow and fast variables, where the latter are averaged in time on the scale of $\Omega^{-1}$ via the Favre procedure [29]. Considering a hydrogen atom, for example, the laminar velocity component $V = 0$ is zero, because the stationary wave function is real and $S = 0$. Therefore, in contrast to the Bohr model the electron does not perform directional motion, but it is simply trembling stochastically, and its trajectory draws the electron cloud, while the kinetic energy is purely turbulent.

The Madelung hydrodynamics (2.1) describes solely the evolution in the coordinate subspace, which follows from the Wigner-Liouville equation in the phase space, similar to the genealogic relationship between Eq. (1.3) and Eq. (1.4). Therefore, we are going now to discover what stochastic dynamics is hidden behind the more general Wigner-Liouville equation [33, 34]. For the sake of simplicity, let us consider the hydrogen atom first, constructed from an electron and a proton with masses $m_1$ and $m_2$, respectively, but the conclusions will be generalized later. The corresponding Schrödinger equation $i\hbar \partial_t \psi = -\hbar^2 \partial_{x_1}^2 \psi / 2m_1 - \hbar^2 \partial_{x_2}^2 \psi / 2m_2 + u_{12} \psi$ describes the evolution of the wave function $\psi(x_1, x_2, t)$ at any positions $x_1$ and $x_2$ of the electron and proton, respectively, where $u_{12} \equiv -e^2 / 4\pi\varepsilon_0 |x_1 - x_2|$ is the Coulomb interaction potential with the elementary charge $e$ and the dielectric permittivity $\varepsilon_0$ of vacuum. The latter shows that vacuum is not empty. Because there is no wave function in classical mechanics, it is appropriate to employ the Wigner function $W(p_1, x_1, p_2, x_2, t)$ as the quasi-probability density in the particles phase space, where $p_1$ and $p_2$ are the electron and proton momenta, respectively. Via the Wigner-Weyl quantization one can calculate all statistical properties in quantum mechanics. The Wigner function is a useful tool and its marginal probability densities coincide with the exact solutions of the Schrödinger equation both in coordinate and momentum representations. The evolution of the Wigner function follows directly from the Schrödinger equation and obeys the Wigner-Liouville equation. For the sake of the further analysis, it is important to express the interaction potential $u_{12}$ via its Fourier image and to write the Wigner-Liouville equation in a decisive form [34]

$$\partial_t W + p_1 \cdot \partial_{x_1} W / m_1 + p_2 \cdot \partial_{x_2} W / m_2 = \int_{-\infty}^{\infty} \frac{e^2 \exp[ik \cdot (x_1 - x_2)]}{\varepsilon_0 i\hbar k^2}$$
$$[W(p_1 + \hbar k/2, x_1, p_2 - \hbar k/2, x_2, t) - W(p_1 - \hbar k/2, x_1, p_2 + \hbar k/2, x_2, t)] d^3(k/2\pi) \quad (2.2)$$

Suddenly, the physics behind the Schrödinger equation becomes transparent. The large expression in the brackets describes transfer of momentum $\hbar k$ from the proton to the electron by a single photon. The other term in the collision integral of Eq. (2.2) is the acting force per unit momentum transmitted, i.e. it is the characteristic frequency of the photon exchange. The ratio between the collision and photon frequencies is the fine structure constant. The integration over positive and negative wave vectors $k$ accounts for momenta transfers in both directions. The photon momentum $\hbar k$ is the only quantum quantity in Eq. (2.2), while the point charges themselves are not quantum and do not propagate as waves. They simply swim in the quantum sea of the photons. This realistic picture correlates well to quantum field theory, stating that the fundamental forces are transmitted by quantum force carriers. Once the non-quantum free particles are set in the force carriers' bath, they become naturally quantum, due to the interaction. One can easily recognize a virtual photon Feynman propagator in the term $-1/k^2$ of Eq. (2.2). The latter seems not quantum, since the photon mass is zero, and it is stationary, because the retardation is neglected in the Schrödinger equation. Obviously, photon is a wave in the coordinate space but it behaves as a quasi-particle in the momentum space, as discovered by Einstein. Therefore, the force carriers are the reason for the wavy character of quantum mechanics. Because the Feynman propagator governs the force carriers of all fundamental interactions, Eq. (2.2) can be easily generalized to describe gravity and strong forces as well. This is the reason why all fundamental interaction potentials vanish inversely proportional to the distance between the point particles. In general, the force carriers transmit either energy or momentum, which creates the interaction potentials and generates the quantum trembling motion as well.

Since many years, the transition of the Wigner-Liouville equation to the classical Liouville equation proves the generic bond between quantum and classical mechanics. If the momenta of the point particles are much larger than the force carriers' quanta, one can expand in a $\hbar k$-series the expression in the brackets in Eq. (2.2). Keeping the leading terms only yields the classical Liouville equation. So, the importance of quantum effects depends on the momenta of the point particles, not on their masses only. Of course, heavy particles possess large momenta even at slow velocity and that is why they obey often classical mechanics. It is evident that a free electron should obey the classical Liouville equation without any external potential since there are not force carriers without interactions. Therefore, the free electron trajectory $X = X_0 + P_0 t/m$ increases linearly in time. However, to fix the electron at its initial position, one must apply a strong initial attractive potential at $X_0$. According to Eq. (2.2) this initial potential will generate momenta kicks $\pm \hbar k/2$, randomizing the electron momentum. Once the initial potential is switched off at $t=0$, the electron will propagate freely with an initial momentum acquired by the photon kicks with $<P_0>=0$. Since the initial variables $<X_0 P_0>=0$ are not correlated, the evolution of the free electron position dispersion $\sigma_x^2 = \sigma_x^2(0) + \sigma_p^2(0) t^2/m^2$ will grow quadratic in time. Due to the properties of the Fourier transform, the initial root-mean-square momentum fluctuation $\sigma_p(0) = \hbar \sigma_k(0)/2 = \hbar/2\sigma_x(0)$ is determined by the width $\sigma_x(0)$ of the initial potential, which emphasizes the measurement problem of quantum mechanics as well. Thus, the Wigner function for a free electron is a Gaussian probability density, which is spreading via the well-known law $\sigma_x^2 = \sigma_x^2(0) + (\hbar t/2m)^2/\sigma_x^2(0)$. The Heisenberg inequality holds at any time, due to of the Fourier transformation properties. The root-mean-square displacement $\sigma_x$ obeys the dynamic equation $m\ddot{\sigma}_x = \hbar^2/4m\sigma_x^3$, which follows from Eq. (2.1) as well. It was introduced in mathematics first by Ermakov in 1880.

Because the fundamental interactions are pairwise additive, the total potential energy is a sum $U = \sum u_{ij}$ of pair potentials. Thus, the many-particles problem reduces straightforward to the same physical picture, emerging from discrete force carriers. Let us consider now a general system of N particles with 3N-dimensional vectors of positions $x$ and momenta $p$, respectively. The Fourier image $\tilde{U}(k)$ of the interaction potential defines the total propagator of the force carriers and represents the interaction energy density as well, distributed over the force carriers' momenta. As was mentioned, the Wigner-Liouville equation describes the evolution of Wigner function, which is traditionally written $\partial_t W + p \cdot M^{-1} \cdot \partial_x W = 2U_{\text{Im}}(x + i\hbar \partial_p / 2)W / \hbar$ by an operator of the potential energy. It can be rewritten, however, in an alternative form, symbolizing the collisions from Eq. (2.2): $\partial_t W + p \cdot M^{-1} \cdot \partial_x W = 2W_{\text{Im}}(p + i\hbar \partial_x / 2, x, t)U / \hbar$. Expressing now the potential $U$ via the corresponding Fourier integral yields after some rearrangements a classically looking alternative

$$\partial_t W + p \cdot M^{-1} \cdot \partial_x W = \int_{-\infty}^{\infty} \tilde{U} \partial_x \exp(ik \cdot x) \cdot \partial_p W_k d^{3N}(k/2\pi) \qquad W_k \equiv \int_{-1/2}^{1/2} W(p + \xi\hbar k, x, t) d\xi \qquad (2.3)$$

As expected, all quantum effects are due to the force carriers' momenta $\hbar k$, randomizing the momenta of the point particles. The normalized probability density $W_k$ offers an alternative interpretation as a random phase approximation of the force carriers' momenta kicks on the point particles. The Bayesian product $\tilde{U}(k)W_k(p, r, t)$ represents the joint distribution of the interaction energy over the force carriers' momenta in the point particles phase space. Again, when the particles momenta $p \gg \hbar k / 2$ are larger than the quanta of the force carriers then $W_k \approx W$ and Eq. (2.3) reduces to the classical Liouville equation $\partial_t W + p \cdot M^{-1} \cdot \partial_x W = \partial_x U \cdot \partial_p W$ but the accuracy of this asymptote depends strongly on the interaction energy Fourier distribution. The unexpected validity of the Liouville equation for harmonic oscillators is due to the fact that the force carriers are also harmonic vibrations. Hence, there is no mode-mode coupling due to linearity of the underlying dynamics. As in the case of a free particle, the quantum effects of a harmonic oscillator originate solely from the initial distribution. For instance, the displacement of a harmonic oscillator with an own frequency $\omega_0$ reads $X = X_0 \cos(\omega_0 t) + P_0 \sin(\omega_0 t) / m\omega_0$ and, if the initial displacement and momentum are not correlated, the position dispersion obeys the expression $\sigma_x^2 = \sigma_x^2(0)\cos^2(\omega_0 t) + \sigma_p^2(0)\sin^2(\omega_0 t) / m^2\omega_0^2$. The initial momentum fluctuation can be expressed again from the initial potential $\sigma_p(0) = \hbar\sigma_k(0)/2 = \hbar/2\sigma_x(0)$. The corresponding position dispersion reduces to the well-known expression $\sigma_x^2 = \hbar/2m\omega_0$, if one is looking for the stationary solution. The oscillator energy is injected by the initial potential during the initial positioning. In contrast to a free particle, $\varepsilon = \hbar\omega_0 / 2$ is universal due to the stationary harmonic oscillator. Because of the parametric resonance, the oscillator can absorb and emit photons with frequency $\omega_0$ only and the possible oscillator energies are $\varepsilon_n = (n+1/2)\hbar\omega_0$. Hence, there is no way to lose the zero-point energy $\hbar\omega_0/2$.

An extensive discussion goes in the literature about the Wigner function positivity. In general, $W$ must possess negative values at some points due to the orthogonality of the stationary solutions of the Schrödinger equation. For the hydrogen atom $W \geq 0$ is positive in the ground state, as well as for a harmonic oscillator. Our expectation is that the Wigner function must be always non-negatively defined in the ground state, while in the excited states $W$ can be negative somewhere. The reason for this is that the excited states appear only at presence of external photons. Hence, the latter will exercise additional momenta kicks on the electron, which are not included in Eq. (2.2) so far. As was demonstrated, the Schrödinger equation is generated by the exchange of force carriers between the interacting particles. Since there are few gauge bosons, Eq. (2.3) is appropriate only for systems with fundamental interactions and the total potential energy must be a superposition of the Feynman momenta propagators. If one employs artificial or even approximate potentials, they will correspond to unphysical propagators of non-existing force carriers, which could result in inconsistent solutions of the Schrödinger equation. It seems that a negative value of the Wigner function is a smart indicator for problems with potentials and initial or boundary conditions. For example, the so-called cat state of a single particle possesses a bimodal Wigner function. Its negativity problem is due, however, to unphysical initial conditions since it is impossible to fix at the beginning a single point particle at two different places at once.

It is interesting how the point particles are moving after all [33]. Borrowing ideas from classical electrodynamics, the Lagrangian $L = \dot{X} \cdot M \cdot \dot{X}/2 - U(X) + A(X) \cdot \dot{X}$ of the entire subsystem takes the form, where $X$ and $\dot{X}$ are the 3N-dimensional vectors of the real trajectories and velocities of the point particles, respectively. Due to energy conservation, the Lagrangian $L$ does not depend explicitly on time. Apart from the scalar potential $U$, the 3N-dimensional vector potential $A$ accounts for the N force carriers. This is consistent with quantum field theory, where the vector potentials are wave functions of the virtual particles in the Klein-Gordon equation. Because the traditional magnetic forces are neglected in Schrödinger equation due to their relativistic character, $A$ is independent of the point particles' velocities $\dot{X}$. Thus, the point particles momenta $P \equiv \partial_{\dot{X}} L$ consist of a differentiable part $M \cdot \dot{X}$ plus kick momenta $A$ and the local-mean of the latter are the laminar and turbulent velocities. The corresponding Euler-Lagrange equation reads $\dot{P} = \partial_X L = -\partial_X U + \partial_X A \cdot \dot{X}$ and one can recognize in the last term the Langevin fluctuation force pumping momenta via kicks. Introducing the particles momenta $P = M \cdot \dot{X} + A$ in this equation yields the corresponding stochastic Langevin-Lorentz equation, which resembles the Brownian motion from Eq. (1.2), [34]

$$M \cdot \ddot{X} + \dot{X} \cdot \partial_X A = -\partial_X U + \partial_X A \cdot \dot{X} \qquad (2.4)$$

In contrast to the standard dissipation, however, the friction tensor $\partial_X A$ of the dissipative force $\dot{A}$ is stochastic and it possesses zero mean value to avoid any entropy production. The stochasticity in Eq. (2.4) originates from the fluctuations of $A(X)$, being a random function of the particles' configuration, because the evolution of the point particles positions causes instant redistribution of the force carriers. The 3N-dimensional Lorentz force $\partial_X A \cdot \dot{X} - \dot{X} \cdot \partial_X A$ does no work since it is always orthogonal to the velocity $\dot{X}$. Hence, the subsystem energy remains constant and $E \equiv P \cdot \dot{X} - L = \dot{X} \cdot M \cdot \dot{X} / 2 + U$ is explicitly independent of the hidden $A$. To respect the Ehrenfest theorem, the mean value of the stochastic Lorentz force must be zero.

The phase space probability density $f = <\delta(x-X)\delta(p-P)>$ is averaged over the stochastic realizations of the vector potential and positively defined everywhere. Differentiating $f$ on time and substituting the Euler-Lagrange equation yields

$$\partial_t f + p \cdot M^{-1} \cdot \partial_x f = \int_{-\infty}^{\infty} ik \cdot \partial_p <\delta(x-X)\delta(p-P)(\tilde{U} - \tilde{A} \cdot \dot{X})\exp(ik \cdot X)> d^{3N}(k/2\pi) \qquad (2.5)$$

In the derivation of this equation the potentials are expressed via the corresponding Fourier forms and the term $\partial_x \cdot M^{-1} \cdot <\delta(x-X)\delta(p-P)A> = 0$ is set zero, because the force carriers' waves are transverse in average. Following the hint from Eq. (2.2), the random potential fluctuations $\tilde{A} \cdot \dot{X} = \tilde{\xi} \hbar k \cdot \dot{X}$ can be expressed by a global fluctuating parameter $\xi(X)$, being dimensionless and zero-centered. Substituting this expression in Eq. (2.5), the latter changes after simple rearrangements to

$$\partial_t f + p \cdot M^{-1} \cdot \partial_x f = \int_{-\infty}^{\infty} ik \cdot \partial_p <\delta(x-X)\delta(p-P)(\tilde{U} + \tilde{\xi} i\hbar \partial_t)\exp(ik \cdot X)> d^{3N}(k/2\pi) \qquad (2.6)$$

It is known from electrodynamics that the gauge transformation $U \to U - \partial_t S$ and $A \to A + \partial_r S$ does not affect the physical state. For instance, one can easily recognize the origin of the quantum mechanical operators of energy and momentum from $S = i\hbar \ln \psi$. The gauge theory explains how the force carriers transmit the potential interaction $U$ via the vector potential $A$. Appling the gauge transformation to cancel and relocate the fluctuation force in Eq. (2.6) results after employing the useful properties of the Dirac delta function in the alternative equation

$$\partial_t f + p \cdot M^{-1} \cdot \partial_x f = \int_{-\infty}^{\infty} \tilde{U} \partial_x \exp(ik \cdot x) \cdot \partial_p f_k d^{3N}(k/2\pi) \tag{2.7}$$

where $f_k \equiv <\delta(x-X)\delta(p-P+\xi\hbar k)>$ reduces to the probability density $f$ at $k=0$. Apparently, the symmetric zero-centered fluctuations of $\xi(X)$ generate the randomness of quantum mechanics, since in the deterministic case ($\xi \equiv 0$) $f_k = f$ and Eq. (2.7) reduces to the classical Liouville equation. To reproduce Eq. (2.3), one should identify the equivalence $f_k \equiv W_k$ between the probability density and the Wigner function, which corresponds to uniformly distributed $\xi$-fluctuations in range $\pm 1/2$. Therefore, the global parameter $\xi$ is the force carriers' helicity, being the projection of their spin on the momentum, which is an important quantum property of the photon beams.

The presented theory above is a kind of stochastic electrodynamics. As was mentioned, the traditional stochastic electrodynamics models describe the temporal fluctuations of the electromagnetic field caused by the zero-point fluctuations in vacuum. Since the latter result always in energy changes, they are excluded in the present model ($\partial_t L = 0$). To compensate the energy fluctuations, the traditional models consider in addition the Abraham-Lorentz force in the stochastic Newton equation but how we showed already, this results in description of the quantum Brownian motion not of quantum mechanics itself. The zero-point fluctuations in vacuum are not accounted in the Schrödinger equation and they cause additionally the Lamb shift of the hydrogen orbitals. Their effect can be estimated by relevant fluctuations in Eq. (2.7), which will change to $\partial_t f + p \cdot M^{-1} \cdot \partial_x f = \hbar^2 \partial_p \cdot M^{-1} \cdot \partial_x^3 f / 4$ in the case of a free particle. Naturally, the corresponding Schrödinger equation is not the usual one anymore. The stationary solution of this equation $\tilde{f} = \tilde{\rho} \exp(-2p^2/\hbar^2 k^2)/(\pi\hbar^2 k^2/2)^{3/2}$, expressed by the Fourier image, is a Gaussian distribution, where $\rho(x)$ is the stationary distribution density in the coordinate space. One can recognize again the virtual kicks in the momentum fluctuation $\pm \hbar k / 2$.

At present, there are too many interpretations of quantum mechanics, which indicate that the problem is not solved, yet. All of them respect the Schrödinger equation but introduce additional concepts to build a physical reality. Our interpretation relays solely on the Schrödinger equation, which is mathematically transformed to the Wigner-Liouville equation, representing the Schrödinger equation in the Wigner-Weyl quantization. This transformation allows us to see clearly the physics behind the Schrödinger equation without any further assumptions. The orthodox Copenhagen interpretation considers the election as a point charge with a wave function, describing the probability for the electron presence. Only the latter exhibits wavy behavior and we have shown in the present paper that the force carriers are causing the probability undulations. The relationship between our and Copenhagen interpretations is like that between statistical mechanics and thermodynamics. Thus, the macroscopic ontology needs our interpretation for the microscopic origin of the Schrödinger equation to recognize its origin. Another important aspect of quantum mechanics is the measurement theory. Because the Schrödinger equation is fundamental, any measurement should affect the system only via changes of the interaction potential. This will change the wave function as well and the traditional theories try to describe this collapse without detail knowledge for the potential disturbances. Our interpretation requires exact specification of the potential during the measurement process. For instance, it was demonstrated that during the fixing of an electron initially the relevant initial potential generates random kicks via its force carriers. These initial kicks are later responsible for the entire quantum behavior of the freed electron. Finally, it appears that photons are the most unique objects, being quantum and moving with the speed of light. Obviously, they are relativistic and do not obey the Schrödinger equation. Remarkably, the vector potential $A$ represents either the electromagnetic field or the wave function of the photon. Hence, the photon is obviously a wave and the d'Alembert equation $\Box A = 0$ is both the classical and quantum governing one. However, because there is no mode-mode coupling in the Fourier image of the wave equation, the photons behave as a perfect gas of quasi-particles in the momentum space. That is how Einstein had explained the discrete photoelectric effect, bringing him the Nobel Prize in 1921. Apparently, the force carriers are the de Broglie pilot waves as the Bohm hidden variables.

## 3. Brownian Motion of Quantum Particles

In 1952, Bohm has developed further the ideas of de Broglie and Madelung by suggesting that quantum particles obey the Newton equation under the action, however, of the quantum potential. Introducing the guiding equation $\dot{X} = V(X,t)$ in Eq. (2.1), the corresponding Newton-Bohm equation $M \cdot \ddot{X} = -\partial_X (U+Q)$ becomes density functional, because the quantum potential $Q \equiv -\hbar^2 \rho^{-1/2} \partial_x \cdot M^{-1} \cdot \partial_x \rho^{1/2} / 2$ depends nonlinearly on the probability density in the coordinate subspace. When $\rho$ is known, the solution of the Newton-Bohm equation is the quantum trajectory of all particles of the closed quantum subsystem. Thus, one can visualize all electron transfers in chemistry, for instance. The Bohmian mechanics is a modern alternative of the orthodox Copenhagen interpretation of quantum mechanics. Because our goal is to describe the Brownian motion of the quantum subsystem, combining the ideas of Bohm and Langevin, we introduced the following stochastic Bohm-Langevin equation [15]

$$M \cdot \ddot{X} + B(X) \cdot \dot{X} = -\partial_X (U + \Phi + Q) + [2k_B T B(X)]^{1/2} \cdot F(t) \tag{3.1}$$

Due to the quantum potential, Eq. (3.1) is a density functional stochastic equation, and it is genetically coupled to the corresponding probability density evolution. Because we know already how to describe the latter, the combination of Eq. (1.4) and Eq. (2.1) yields the quantum hydrodynamic-like equations, resembling Eq. (3.1), [14, 33]

$$\partial_t \rho + \partial_x \cdot (\rho V) = 0 \qquad M \cdot (\partial_t V + V \cdot \partial_x V) + B \cdot V = -\partial_x (U + \Phi + Q + k_B T \ln \rho) \tag{3.2}$$

This system of equations is self-consistent and provides simultaneously the probability density $\rho$ and the hydrodynamic-like velocity $V$ as its solutions. In the case of a strong friction, one can neglect the inertial term in Eq. (3.2) to express the hydrodynamic-like velocity via the probability density. Substituting the following $V = -B^{-1} \cdot \partial_x (U + \Phi + Q + k_B T \ln \rho)$ in the continuity equation yields the Smoluchowski-Bohm equation [15]

$$\partial_t \rho = \partial_x \cdot B^{-1} \cdot [\rho \partial_x (U + \Phi + Q) + k_B T \partial_x \rho] = \partial_x \cdot B^{-1} \cdot [\rho \partial_x (\rho^{-1/2} \hat{H} \rho^{1/2}) + k_B T \partial_x \rho] \tag{3.3}$$

where $\hat{H}$ is the subsystem Hamiltonian operator in coordinate representation. Linearizing Eq. (3.3) results in a quantum Smoluchowski equation with the position-acting temperature operator $k_B\hat{T} = k_BT - \hbar^2 \partial_x \cdot M^{-1} \cdot \partial_x / 4$, which reflects the quantum nature of the subsystem particles now [16, 19]. The corresponding osmotic pressure $p = k_B\hat{T}\rho$ is naturally a superposition of the already mentioned thermal and Madelung quantum components.

The Smoluchowski-Bohm equation is a closed mathematical problem for the evolution of the probability density in the coordinate subspace. Unfortunately, the equilibrium solution of Eq. (3.3) is not the exact equilibrium quantum distribution known from statistical mechanics, but the reason is obvious. Because the probability density ρ depends on temperature as well the thermal fluctuations are accounted twice in Eq. (3.1). To correct this non-additivity, one could replace the quantum potential by the corresponding free energy $k_BT\int Qd\beta$, calculated by the Gibbs-Helmholtz relation. Thus, the Smoluchowski-Bohm equation acquires the enhanced form [7, 10]

$$\partial_t\rho = \partial_x \cdot k_BTB^{-1} \cdot [\rho\partial_x\int_0^\beta \rho^{-1/2}(\hat{H} + 2\partial_\beta)\rho^{1/2}d\beta] \tag{3.4}$$

One can recognize the Einstein diffusion tensor in $k_BTB^{-1}$, while according to the non-equilibrium thermodynamics the integral must represent the non-equilibrium free energy functional in the coordinate subspace, divided by $k_BT$. Since the latter equals to $-\ln Z$ at equilibrium, where $Z$ is the quantum partition function, the equilibrium solution of Eq. (3.4) obeys the Bloch type equation $2\partial_\beta(\rho_{eq}Z)^{1/2} = -\hat{H}(\rho_{eq}Z)^{1/2}$. Its solutions $\rho_n(x) = \exp(-\beta E_n)\varphi_n^2(x)/Z$ are the quantum canonical Gibbs distributions, where $E_n$ and $\varphi_n$ are the eigenvalues and normalized eigenfunctions of the subsystem Hamiltonian, respectively. The well-known expression for the quantum partition function $Z = \sum\exp(-\beta E_n)$ follows from normalization. As is seen, the integral in Eq. (3.4) leads to the correct equilibrium distribution but it complicates additionally the nonlinear mathematical problem. Fortunately, either at low or at high temperature Eq. (3.4) reduces to Eq. (3.3), which can be proven by the use of the l'Hopital rule. For this reason, in the further applications we will focus on Eq. (3.2). Considering, for instance, an adsorbed particle with $\Phi = m\omega_0^2x^2/2$, the friction coefficient is constant, and the probability density is Gaussian. Hence, the resulting

Ermakov equation $m\ddot{\sigma}_x + b\dot{\sigma}_x + m\omega_0^2\sigma_x = \hbar^2/4m\sigma_x^3 + k_BT/\sigma_x$ follows from Eq. (3.2) and accounts for many particular cases. It reproduces the exact quantum equation at zero temperature and zero friction coefficient. At high friction one can neglect the first inertial term to obtain an equation for the dispersion evolution $b\partial_t\sigma_x^2 + 2m\omega_0^2\sigma_x^2 = \hbar^2/2m\sigma_x^2 + 2k_BT$, which follows from Eq. (3.3) as well. In the classical limit it predicts $\sigma_x^2 = (k_BT/m\omega_0^2)[1-\exp(-2m\omega_0^2 t/b)]$, which is the well-known solution of Eq. (1.5). In the opposite case at zero temperature the position dispersion relaxes quicker than the classical one as $\sigma_x^2 = (\hbar/2m\omega_0)\sqrt{1-\exp(-4m\omega_0^2 t/b)}$.

The Smoluchowski-Bohm equation describes the evolution in the coordinate subspace only, but the complete mechanical description requires master equation acting in the subsystem phase space. Due to mathematical complications, we will consider further structureless environments via spatially averaged friction tensor $\bar{B} = \bar{b}I$ with position independent friction coefficient $\bar{b}$ and potential $\bar{\Phi} = 0$. It is possible to quantize the classical Klein-Kramers equation directly by replacing the canonical derivatives and functional products via the commutators [,] and anti-commutators {,}, respectively. By this technique Eq. (1.3) transforms to the Caldeira-Leggett equation $\partial_t\hat{\rho} - [\hat{H},\hat{\rho}]/i\hbar = \bar{b}[\hat{x},\{\hat{\rho},[\hat{x},\hat{H}]/i\hbar\}/2 + k_BT[\hat{x},\hat{\rho}]/i\hbar]/i\hbar$ for the density operator $\hat{\rho}$ of the N-particles subsystem, which reduces to the Liouville-von Neumann equation for the closed subsystem at $\bar{b} = 0$. Conventionally, the superscript as that in the Hamiltonian $\hat{H}$ denotes quantum mechanical operators in the Heisenberg picture. It is well known that the Caldeira-Leggett equation is correct only at high temperature and that is why its equilibrium solution differs from the rigorous quantum canonical Gibbs density operator $\hat{\rho}_{eq} = \exp(-\beta\hat{H})/Z$. Introducing the Wigner function $W$, being the quantum analog of the classical phase space probability density $f$, the Caldeira-Leggett equation can be straightforward transformed to

$$\partial_t W - 2H\sin\vec{\Lambda} W/\hbar = \bar{b}\partial_p \cdot (W\cos\vec{\Lambda}\partial_p H + k_BT\partial_p W) \qquad (3.5)$$

The arrows in the super operator $\vec{\Lambda} \equiv \hbar(\vec{\partial}_x \cdot \vec{\partial}_p - \vec{\partial}_p \cdot \vec{\partial}_x)/2$ indicate the direction of differentiation and the commutators and anti-commutators change to $2i\sin\vec{\Lambda}$ and $2\cos\vec{\Lambda}$, respectively. Since Eq. (3.5) reduces to the Wigner-Liouville equation in the case $\bar{b} = 0$, it accounts rigorously for quantum mechanics on the left-hand side but the last diffusional term on the right-hand side is purely classical. This semiclassical discrepancy results in an approximate equilibrium solution.

For instance, Eq. (3.5) reduces exactly to the classical Eq. (1.5) in the case of harmonic oscillators with the Hamilton function $H \equiv p^2/2m + m\omega_0^2 x^2/2$. Thus, any initial quantum correlation will disappear during the irreversible evolution and the quantum oscillators will become classical at equilibrium. Traditionally, this crucial problem is fixed by replacing the thermal energy $k_B T$ via the mean energy $\varepsilon = (\hbar\omega_0/2)\coth(\beta\hbar\omega_0/2)$ of a quantum Brownian oscillator at equilibrium to obtain

$$\partial_t W + p \cdot \partial_x W/m - m\omega_0^2 x \cdot \partial_p W = \bar{b}\partial_p \cdot [pW/m + (\hbar\omega_0/2)\coth(\beta\hbar\omega_0/2)\partial_p W] \qquad (3.6)$$

Such approach is, however, neither rigorous nor universal and demonstrates again the thermodynamic limitations of the Caldeira-Leggett equation. Enhancement of the latter to the Lindblad form fails also to reproduce the exact equilibrium density operator in general.

The main goal of the third part of the Dissertation is to improve the widely used Caldeira-Leggett equation [35]. Strictly speaking, the Markov processes exist neither in classical nor in quantum mechanics, but they are the most reliable and simple idealizations for any dynamics. As was mentioned, the rigorous approach requires integration of the exact Liouville-von Neumann equation, which is possible for harmonic oscillators only, because of the linearity of the corresponding dynamic equations. In this way the Markov Caldeira-Leggett equation is derived by employing some additional hypotheses, e.g. the factorization of the initial quantum state in the Feynman-Vernon approach from 1963. For this reason, we will try also to map the quantum dynamics on the Markov one. According to the Onsager non-equilibrium thermodynamics, the flow is proportional to the gradient of the relevant thermodynamic potential, which is the non-equilibrium local free energy functional $F \equiv H + k_B T \ln f$. Respecting this deeper physics, one should rewrite Eq. (1.5) in the more general form $\partial_t f + \partial_p H \cdot \partial_x f - \partial_x H \cdot \partial_p f = \bar{b}\partial_p \cdot (f\partial_p F)$. Quantizing now the latter yields

$$\partial_t \hat{\rho} - [\hat{H},\hat{\rho}]/i\hbar = \bar{b}[\hat{x},\{\hat{\rho},[\hat{x},\hat{H}+k_B T\ln\hat{\rho}]/i\hbar\}/2]/i\hbar \qquad (3.7)$$

It is obvious that the Gibbs density matrix is the equilibrium solution of Eq. (3.7). A fundamental difference between this master equation and the Caldeira-Leggett equation is the Boltzmann logarithm, originating from the subsystem entropy. The classical Klein-Kramers equation is linear due to the differentiation of the entropy, while Eq. (3.7) remains nonlinear, owing to the non-commutative quantum algebra. It is known that the exact von Neumann entropy $-k_B tr(\hat{\rho}\ln\hat{\rho})$ differs from the Shannon-Wigner entropy $-k_B \int W \ln W dp dx$ [25], which is driving the classical diffusion in Eq. (3.5), although the energy $E \equiv tr(\hat{\rho}\hat{H}) = \int HW dp dx$ is the same in both representations. The nonlinearity of Eq. (3.7) changes dramatically the quantum evolution of open systems by repealing the superposition principle. This requires a critical reassessment of the quantum decoherence, described traditionally via linear master equations.

One can linearize Eq. (3.7) around the exact equilibrium density operator $\hat{\rho}_{eq}$ to obtain $\partial_t \hat{\rho} - [\hat{H},\hat{\rho}]/i\hbar = \bar{b}k_B T[\hat{x},\{\exp(-\beta\hat{H}),[\hat{x},\{\exp(\beta\hat{H}),\hat{\rho}\}/2]/i\hbar\}/2]/i\hbar$ and $\hat{\rho}_{eq} = \exp(-\beta\hat{H})/Z$ is naturally the equilibrium solution of this equation. If one considers further the high temperature limit and linearizes the exponential operators as well, this equation reduces to the Caldeira-Leggett equation, as expected. An advantage of the linearity is that it can be directly transformed in the Wigner phase space

$$\partial_t W - 2H\sin\vec{\Lambda}W/\hbar = \bar{b}k_B T \partial_p \cdot \{\exp(-\beta H\cos\vec{\Lambda})\partial_p[\exp(\beta H\cos\vec{\Lambda})W]\} \qquad (3.8)$$

As is seem, the formal equilibrium solution $W_{eq} = \exp(-\beta H\cos\vec{\Lambda})/Z$ obeys the Bloch-Wigner equation $\partial_\beta(W_{eq}Z) = -H\cos\vec{\Lambda}W_{eq}Z$, as required. In the simplest case of an ideal gas, the Hamilton function $H \equiv p^2/2m$ depends on the momenta of the subsystem particles only and Eq. (3.5) coincides with the classical Klein-Kramers equation. Surprisingly, Eq. (3.8) reduces also to Eq. (1.3), which shows that quantum effects for free Brownian particles must be nonlinear. For harmonic oscillators, the super operator $H\cos\vec{\Lambda} = H - H\vec{\Lambda}^2/2$ splits to two parts, depending on $p$ and $x$, respectively. The contributions of the $x$-part cancel in relaxation term of Eq. (3.8) since it commutes with $\partial_p$. Because the second derivative on $\beta$ of the relaxation operator for the Brownian harmonic oscillators equals to the operator itself multiplied by $(\hbar\omega_0/2)^2$, the latter is a linear combination of the hyperbolic sine and cosine functions of $\beta\hbar\omega_0/2$. Therefore, Eq. (3.8) acquires the following particular form

$$\partial_t W + p \cdot \partial_x W / m - m\omega_0^2 x \cdot \partial_p W = b\partial_p \cdot [pW/m + (\hbar\omega_0/2)\coth(\beta\hbar\omega_0/2)\partial_p W] \qquad (3.9)$$

Both Eq. (3.6) and Eq. (3.9) are linear and possess the same exact equilibrium solution but $W_{eq}$ is derived from Eq. (3.9) and presumed in Eq. (3.6). The quantum effect in Eq. (3.6) is solely prescribed to diffusion, while in Eq. (3.9) both the friction and diffusion are quantum. The emergent friction coefficient $b \equiv \bar{b}\sinh(\beta\hbar\omega_0/2)/(\beta\hbar\omega_0/2)$ agrees with the Wigner quantum transition state theory at zero barrier, since $\sinh(\beta\hbar\omega_0/2)$ is inversely proportional to the partition function of a quantum oscillator. The momentum diffusion coefficient $D_p \equiv \bar{b}k_B T \cosh(\beta\hbar\omega_0/2)$ is also amplified but obeys the fluctuation dissipation theorem $D_p = b(\hbar\omega_0/2)\coth(\beta\hbar\omega_0/2)$ for quantum systems. The adiabatic friction coefficient $\beta D_p$ is larger than the isothermal $b$ and they are related via the Gibbs-Helmholtz equation $\partial_\beta (\beta b)_{\bar{b}} = \beta D_p$ [1]. At zero temperature, the friction coefficients diverge, because $\hbar\omega_0/2$ plays the role of activation energy as well, and the harmonic oscillator drops at once in the equilibrium ground state with the well-known Wigner function $W_{eq} = \exp(-2H/\hbar\omega_0)/Z$. The quantum oscillator moves in the grounds state without any friction due to the tunneling effect, but to move macroscopically it should be exited first. At zero temperature the environment cannot supply the necessary excitation energy $\hbar\omega_0$, which reflects in the infinite emergent friction coefficient $b$. This effect weakens, however, by a decrease of the collision frequency $\bar{b}/m$, which at zero temperature is solely due to the quantum motion of the subsystem particles in the ground state [21, 31].

Formally, it is possible to convert Eq. (3.7) in the Wigner representation

$$\partial_t W - 2H \sin \vec{\Lambda} W / \hbar = \bar{b}\partial_p \cdot \{W\partial_p[\cos \vec{\Lambda} H + k_B T \ln(\cos \vec{\Lambda} W)]\} \qquad (3.10)$$

Using the operator equality $\cos \vec{\Lambda} \exp(-\beta H \cos \vec{\Lambda}) = \exp(-\beta \cos \vec{\Lambda} H)\cos \vec{\Lambda}$ one can prove that the equilibrium solution of Eq. (3.10) is the exact $W_{eq} = \exp(-\beta H \cos \vec{\Lambda})/Z$ again. Extracting the Shannon-Wigner entropy, Eq. (3.10) can be further presented in the form of Eq. (3.5). It is evident now that the nonlinear operator $-k_B \ln(\cos \vec{\Lambda} W / W)$ represents the quantum entropy, vanishing naturally in the classical limit $\hbar \to 0$. It persists even at zero temperature to ensure the correct quantum distribution in the ground state. Solving the nonlinear Eq. (3.10) in general is a mathematical problem more difficult than quantum mechanics of closed systems, because the Wigner-Liouville part is much simpler than the relaxation one. However, taking the leading quantum corrections, $\sin \vec{\Lambda} \approx \vec{\Lambda} - \vec{\Lambda}^3 / 6$ and $\cos \vec{\Lambda} \approx 1 - \vec{\Lambda}^2 / 2$, and expanding the logarithm in series as well yield a semiclassical Klein-Kramers equation

$$\partial_t W - 2H\vec{\Lambda}W/\hbar + H\vec{\Lambda}^3 W/3\hbar = \bar{b}\partial_p \cdot [W\partial_p H + k_B T \partial_p W - k_B T W \partial_p (\vec{\Lambda}^2 W / 2W)] \tag{3.11}$$

The last linear quantum term on the left-hand side is well known and vanishes for free particles and oscillators. The quantum term on the right-hand side is nonlinear. It accounts for the Fisher entropy via the nonlinear Bohm quantum potential represented in the Wigner phase space [16]. For numerical applications in chemistry, for instance, a TDDFT image of Eq. (3.10) is already proposed via a nonlinear dissipative thermo-quantum Kohn-Sham equation [17].

Let us return to the harmonic oscillators. Although the relevant Eq. (3.11) is nonlinear, its solution is a normal distribution. Using the bivariate Gaussian Wigner function for each oscillator, the nonlinear quantum term acquires the linear form $k_B T \hbar^2 \partial_p W / 4(\sigma_x^2 \sigma_p^2 - \sigma_{xp}^2)$. The relaxation effect is important at large friction, where the Brownian motion of the subsystem particles becomes overdamped at $t > \tau_1 \equiv m/\bar{b}$. In this case, the fast thermalization in the momentum subspace is already over and the observation follows solely the slow relaxation in the coordinate subspace. Because the nonlinear term is a quantum correction, one should employ therein the relevant classical expressions for the momentum dispersion $\sigma_p^2 = mk_B T$ and correlation $\sigma_{xp} = 0$ at equilibrium. Hence, substituting the Bohmian term $\hbar^2 \partial_p W / 4m\sigma_x^2$ back in Eq. (3.11) yields an emergent Fokker-Planck equation

$$\partial_t W + p \cdot \partial_x W / m - m\omega_0^2 x \cdot \partial_p W = \bar{b}\partial_p \cdot [pW/m + (k_B T + \hbar^2 / 4m\sigma_x^2)\partial_p W] \tag{3.12}$$

One can see immediately that the quantum entropy increases the thermal energy by the Heisenberg momentum uncertainty, i.e. the classical environment monitors continuously the quantum subsystem by measurements. This non-equilibrium thermo-quantum expression substitutes the equilibrium momentum dispersion in Eq. (3.6). Combining the equilibrium Maxwell-Heisenberg relation $\sigma_p^2 = mk_B T + \hbar^2 / 4\sigma_x^2$ with the virial theorem $m\omega_0^2 \sigma_x^2 = \sigma_p^2 / m$ yields the mean energy at equilibrium $\varepsilon = (k_B T / 2)[\sqrt{1+(\beta\hbar\omega_0)^2}+1]$, which is slightly higher than the well-known exact expression $\varepsilon = (\hbar\omega_0 / 2)\coth(\beta\hbar\omega_0 / 2)$ due to the semiclassical approximations in Eq. (3.11) [15]. Both expressions coincide, however, at zero and infinite temperature. Following the standard procedure at large friction coefficient $\bar{b}$ one can derive from Eq. (3.12) the Smoluchowski-Bohm equation (3.3) for the harmonic oscillators

$$\partial_t \rho = \partial_x \cdot [m\omega_0^2 x\rho + (k_B T + \hbar^2/4m\sigma_x^2)\partial_x \rho]/\bar{b} = \partial_x \cdot [\rho\partial_x(U+Q)/\bar{b} + D\partial_x \rho] \tag{3.13}$$

Again, the Smoluchowski-Bohm equation, governing the probability density in the coordinate subspace, does not provide the exact equilibrium distribution in general, because of the semiclassical approximations in the quantum entropy, but this can be improved by Eq. (3.4).

Finally, let us reconsider the most interesting case of an ideal gas by setting $\omega_0 \equiv 0$. As discussed before, in this case Eq. (3.6) becomes classical. Now the Maxwell-Heisenberg relation provides the exact value at equilibrium, because $\sigma_x^2$ diverges in time. For free particles, Eq. (3.13) reduces to diffusion equation, where the Planck constant scales to the thermal de Broglie wavelength $\lambda_T \equiv \hbar/2\sqrt{mk_B T}$ in the dispersion-dependent diffusion coefficient $D(1+\lambda_T^2/\sigma_x^2)$. The direct integration of the standard diffusional equation $\partial_t \sigma_x^2 = 2D(1+\lambda_T^2/\sigma_x^2)$ unveils the quantum generalization of the classical Einstein law of Brownian motion [13, 14]

$$\sigma_x^2 - \lambda_T^2 \ln(1+\sigma_x^2/\lambda_T^2) = 2Dt \tag{3.14}$$

The quantum relaxation time $\tau_2 \equiv \lambda_T^2/2D = \bar{b}/2m\omega_2^2$ corresponds to an oscillator with the second Matsubara frequency $\omega_2 \equiv 2k_B T/\hbar$. The classical Einstein law $\sigma_x^2 = 2Dt$ holds when $t > \tau_2$, which is easily achieved at high temperature, but at short time a purely quantum expression follows from Eq. (3.14). This sub-diffusive quantum law $\sigma_x^2 = \hbar\sqrt{t/m\bar{b}}$ is our central invention, being always valid at low temperature, where the quantum entropy dominates over the classical one [19]. Because Eq. (3.14) is derived for large time, the necessary condition to be able to see quantum effects is $\tau_2 > \tau_1$, i.e. the Nelson diffusion constant $\hbar/2m > D$ must be larger than the Einstein one, which is typical for light particles at low temperature and high friction. In the case of the periodic Frankel-Kantorova model the purely quantum diffusion at zero temperature, described via the Festa-d'Agliano formula, becomes logarithmic $\sigma_x^2 = (\hbar^2/8mA)\ln(32\pi mA^2 t/\bar{b}\hbar^2)$ and the position dispersion is proportional to the de Broglie wavelength of the activation energy [19].

The Planck constant appears in the present part of the Dissertation solely from the subsystem quantum operators. Therefore, the considered thermal bath is classical and affects the subsystem particles only via the friction constant $b$ and temperature $T$. For this reason, the Smoluchowski-Bohm equation describes classical diffusion in the fields of classical and quantum potentials. A new discrete model for the energy relaxation of a quantum particle in a classical environment is proposed via a projection operator $\hat{P}\psi = |\psi|$, causing the wave function collapse [31]. Using this original operator, which preserves the probability density, various power laws for the evolution of the coordinate and momentum dispersions of the quantum Brownian particle are derived. New dissipative Schrödinger and Liouville equations are also obtained and solved for particular cases. In general, the environment can be quantum as well, which complicates additionally the theoretical analysis via a time-dependent temperature operator $k_B\hat{T}$ and more complex quantum friction, which can affect the equilibrium distribution as well. Therefore, it is essential to distinguish our Brownian motion of quantum particles in a classical environment from the more complex Brownian motion in a quantum environment. It is well known that $\sigma_x^2$ grows logarithmically in time for the quantum Brownian motion in an environment with non-Markov retardation at zero temperature. Interestingly, this quantum bath effect can also be accounted via the thermo-quantum Maxwell-Heisenberg relation $\sigma_p^2 = mk_BT + \hbar m/t + \hbar^2/4\sigma_x^2$, enhanced by the Heisenberg time-energy uncertainty [21]. The quantum corrections here are solely the first two terms in an infinite series on the powers of the Planck constant. The linear term accounts for the quantum environment, since it supplies pure energy, while the particle quantum contribution is given by the quadratic term, because it goes through the particle momentum. Knowing the influence of potentials in quantum mechanics, we expect a dramatic quantum effect of the position dependent friction coefficient $b(x)$ in structured media, as well as of the nonuniform viscous friction between the subsystem particles [18, 20].

## 4. Conclusions

A. In the frames of classical mechanics, the generalized Langevin equation is derived for an arbitrary mechanical subsystem coupled to the harmonic bath of a solid and the statistical properties of the multiplicative stochastic Langevin force are rigorously obtained. An important expression is derived, which relates the friction tensor to the static subsystem-bath interaction potential and demonstrates nonuniformity of the friction force in structured environments.

B. The relevant classical Langevin equation is properly applied for description of many examples of resonant Brownian motion of atoms, rigid, rotating and vibrating dimers, and normal alkanes in zeolites. In all cases the nonmonotone dependence of the classical diffusion constant on the structure of the diffusing species is confirmed, corresponding quantitatively to several experimental observations.

C. A time-acting temperature operator is introduced for the quantum Klein-Kramers and Smoluchowski equations, accounting for the effect of the quantum thermal bath oscillators on the target motion of the coupled classical subsystem. This emergent idea is extended to living species as well, where the temperature is replaced by a new quantity, called temperament, which reflects the motivity of living cells. The corresponding vital Klein-Kramers and Smoluchowski equations are derived, and several solutions are obtained.

D. The model of Brownian emitters is theoretically studied in details and the corresponding evolutionary equations for the probability density are derived, which possesses friction coefficient operator acting in time. It is shown that the motion of an electron, swimming in the bath of the zero-point vacuum energy fluctuations, is non-Markov Brownian one and it does not lead to the Schrödinger equation as claimed by the theory of stochastic electrodynamics.

E. The fundamental Schrödinger equation is explained as a result of the collisions of the target point particles with the quantum force carriers, transmitting the fundamental interactions between the point particles. Thus, electrons and other point particles are no waves and the wavy chapter of quantum mechanics originated for the force carriers, being waves in the coordinate subspace and quasi-particles in the Fourier momentum subspace.

F. A stochastic Lorentz equation is proposed as a kind of Langevin equation, which described the underlaying Brownian-like motion of the point particles in quantum mechanics. The stochasticity originates from the unknown behavior of the force carriers acting via stochastic vector potentials. It is demonstrated how the Schrödinger equation can be derived from the Lorentz-Langevin equation by using gauge transformation.

G. Considering the Brownian dynamics in the frames of the Bohmian mechanics, the density functional Bohm-Langevin equation is proposed. The relevant Smoluchowski-Bohm equation is derived, which describes the evolution of the probability density of a quantum subsystem, coupled to a classical environment.

H. A nonlinear master equation is proposed by quantization of the proper form of the classical Klein-Kramers equation. Its equilibrium solution in the exact canonical Gibbs density operator, while the well-known Caldeira-Leggett equation is simply a linearization at high temperature.

I. The application of our nonlinear master equation to harmonic oscillators confirms the Smoluchowski-Bohm equation. In the case of a free quantum Brownian particles, a new law for the spreading of the wave packet it discovered, which represents the quantum generalization of the classical Einstein law of Brownian motion.

J. A new projector operator is proposed for the collapse of the wave function of a quantum particle moving in a classical environment. Its application results in dissipative Schrödinger equations, as well as in a new form of dissipative Liouville equation in classical mechanics.